\documentstyle[12pt,epsf,axocolor]{article}
\newcommand{\bara}{\begin{array}{c}}
\newcommand{\eara}{\end{array}}

\def\barr{\begin{array}}
\def\earr{\end{array}}
\def\dis{\displaystyle}
\def\be{\begin{equation}}
\def\ee{\end{equation}}
\def\ba{\begin{eqnarray}}
\def\ea{\end{eqnarray}}
\def\bann{\begin{eqnarray*}}
\def\eann{\end{eqnarray*}}
\def\benn{\begin{displaymath}}
\def\eenn{\end{displaymath}}

\def\etal{ {\em et al.}}
\def\fb{\: {\rm fb}}
\def\gev{\: {\rm GeV} }
\def\tev{\: {\rm TeV} }
\def\pb{\: {\rm pb}}
\def\ra{\rightarrow}
\def\ie{ {\em i.e.}}

\def\lsim{\:\raisebox{-0.5ex}{$\stackrel{\textstyle<}{\sim}$}\:}

\def\mybox#1{\raisebox{-1pt}[0pt][6pt]{\framebox(10,10) [0.5cm]{#1}} }

\def\kt{\tilde{\kappa}}
\def\lt{\tilde{\lambda}}

\begin{document}
\thispagestyle{empty}
\setcounter{page}{0}
 

\begin{flushright}
DTP-99/06\\
IFT/99-01\\
MRI-PHY/P990173\\

{\large \tt hep-ph/9904215}
\end{flushright}

\vskip 45pt
\begin{center}
{\Large \bf  
        {\boldmath $CP$}-violating Anomalous {\boldmath $WW\gamma$} 
          Couplings in {\boldmath $e^+e^-$}~Collisions}

\vspace{10mm}
{\large Debajyoti Choudhury$^1$, Jan Kalinowski$^2$ and Anna Kulesza$^3$}
\end{center}

\vspace{6mm}

\noindent $^1$ Mehta Research Institute, Chhatnag Road, Jhusi,
               Allahabad--211 019, India\\
          \hspace*{0.7em}
             {\sl E-mail: debchou@mri.ernet.in, debchou@mail.cern.ch}
\\[0.8ex]
$^2$ Instytut Fizyki Teoretycznej, Uniwersytet Warszawski, Hoza 69, 00691
Warszawa, Poland\\
          \hspace*{0.7em}
             {\sl E-mail: Jan.Kalinowski@fuw.edu.pl}\\[0.8ex]
$^3$ Department of Physics, University of Durham, Durham DH1 3LE, UK\\
          \hspace*{0.7em}
             {\sl E-mail: anna.kulesza@durham.ac.uk}

\vspace{10mm}

\begin{abstract}
We investigate the sensitivity of future linear collider experiments
to $CP$ violating $WW\gamma$ couplings in the process $e^{+}e^{-} \ra
\nu\bar{\nu}\gamma$. We consider several sets of machine parameters:
centre of mass energies $\sqrt{s}$ = 350, 500 and 800 GeV and
operating at different luminosities. From an analysis of the
differential cross-section the following 95\% C.L. limits
$|\tilde{\kappa}_{\gamma}|<0.18$, $|\tilde{\lambda}_{\gamma}|<0.069$
are estimated to be obtained at a future 500 GeV LC with an integrated
luminosity of $125 \fb^{-1}$, a great improvement as compared
to the LEP2 reach, where a senstitivity of order 2 for both couplings
is found.
\end{abstract}
\vspace{5cm}

\pagebreak
\section{Introduction}
        \label{sec:intro}

The ongoing LEP run at energies above the
$W^+W^-$ threshold has made it possible to study directly the
non-Abelian structure of the electroweak Standard Model (SM) in the
clean environment of $e^+e^-$ collisions.  
LEP2 has not only confirmed the existence 
of  triple gauge-boson vertices, inferred either 
indirectly~\cite{indirect} or from observations at $p\bar{p}$ 
colliders~\cite{tev}, but also established constraints on them~\cite{LEP}.
The goal of the present studies at LEP2, and at future 
colliders, is to test the structure of the bosonic sector with
a precision comparable to that achieved for the fermion-vector boson
couplings. Such precise measurements of gauge vector-boson
couplings will not only provide stringent tests of the gauge structure
of the SM, but also probe for new physics.

Within the Standard Model, triple and quartic vector-boson
interactions are intimately related to the gauge structure of the
model and therefore are completely determined. Of course, radiative
corrections within the SM modify the tree-level couplings. However,
such corrections are quite small. In particular for the $CP$-violating
couplings they are expected to be exteremely small and unmeasurable in
near future since $e.g.$ the electric dipole moment of the $W$ boson
vanishes to two loops\cite{KP}. On the other hand, any theory
incorporating new physics may conceivably induce much larger (already
at the one-loop) deviations in some of the couplings. Corrections at a
permille level can be expected in multi-Higgs or supersymmetric
extensions\cite{hisusy}. Models with dynamical breaking of electroweak
symmetry by new strong forces, could produce even larger corrections
\cite{dynmodel}. The concomitant $CP$ violation could then manifest
itself in non-zero $CP$-violating gauge boson couplings, observation of
which would be a clear signal of beyond-SM physics.

Owing to the general perception that the $CP$-violating couplings are
severely constrained by the data on the neutron electric dipole moment
(EDM)~\cite{edmn}, in phenomenological analyses of the physics
potential of future colliders, their sensitivity to these couplings
has received less attention than that accorded to the $CP$-conserving
ones. However, with these constraints being the subject to naturalness
assumption, there is no substitute for a direct measurement.
Furthermore, since they depend on different combinations of anomalous
couplings, the direct measurements are complementary to the indirect
analyses. In the present paper we wish to study the sensitivity of
future $e^+e^-$ linear colliders (LC) to $CP$-violating couplings, in
particular we will consider the process $e^+e^-\ra \nu\bar{\nu}
\gamma$ as a means to test $WW\gamma$ interactions. The motivation for
our study, of course, is that the origin of $CP$ violation remains
unexplained and it should be experimentally investigated wherever
possible. We find that at an $e^+e^-$ 500 GeV linear collider with an
integrated luminosity of ${\cal L}=125 \fb^{-1}$, an analysis of the
differential cross-section allows us to derive the following 95\% C.L.
limits $|\tilde{\kappa}_{\gamma}|<0.18$,
$|\tilde{\lambda}_{\gamma}|<0.069$. With a high luminosity option of
500 fb$^{-1}$, more stringent limits $|\tilde{\kappa}_{\gamma}|<0.13$,
$|\tilde{\lambda}_{\gamma}|<0.049$ can be established. For comparison,
we find that the LEP2 experiments at $\sqrt{s}=192$ GeV and ${\cal L}=2
\fb^{-1}$ can reach a sensitivity of the order
$|\tilde{\kappa}_{\gamma}| \sim |\tilde{\lambda}_{\gamma}|\sim 2$

\section{Anomalous Couplings}
        \label{sec:anom}
It is convenient to describe the phenomenology at scales well below
the scale of new physics in a model independent way. It is, by now,
standard to introduce an effective low-energy Lagrangian that contains
only SM fields. This assumes that the physics responsible for any
deviations is not directly observable, but can  manifest itself
through virtual corrections.  This formalism provides a simple
parametrization of the triple gauge-boson couplings (TGC).  In purely
phenomenological terms, the effective Lagrangian for the $WW\gamma$
and the $WWZ$ interactions can be expressed in terms of seven
parameters each~\cite{hpzh} {\em viz.}
\be
\barr{rl} \dis
     {\cal L}_{\it eff}^{WWV} = &
      \dis -i g_{V} \Bigg[ \hspace*{0.3em}
                         ( 1 + \Delta g^V_1 )
                           \left( W^\dagger_{\mu \nu} W^\mu
                                 - W^{\dagger\mu} W_{\mu \nu}
                           \right) V^\nu +
                          ( 1 + \Delta \kappa_V)
                            W^\dagger_{\mu} W_\nu V^{\mu\nu}\\[1.5ex]

            & \hspace*{2.3em} \dis
               + \frac{\lambda_V}{M_W^2}
                 W^\dagger_{\mu \nu} {W^\nu}_\sigma
                 V^{\sigma\mu}  -i g^V_5
         \epsilon^{\mu\nu\rho\sigma}(W^\dagger_\mu 
\partial_{\rho} W_{\nu}- W_{\nu}\partial_{\rho}W^\dagger_{\mu}) 
V_\sigma
\\[1.5ex]
            &\hspace*{2.3em} \dis 
             + 
      i g_{4}^{V}W_{\mu}^{\dagger}W_{\nu}(\partial^{\mu}V^{\nu} 
     +\partial^{\nu}V^{\mu})
+ \kt_{V}W_{\mu}^{\dagger}W_{\nu}\tilde{V}^{\mu\nu}+ 
\frac{\lt_{V}}{m_{W}^2}W_{\lambda\mu}^{\dagger}W_{\ \
\nu}^{\mu}\tilde{V}^{\nu\lambda}\;\Bigg{]}\:,
\earr
      \label{lagrangian}
\ee
where $V^{\mu}$ is a neutral vector boson field, i.e. either the $\gamma$
or the $Z$ field. The $W^{\mu}$ ($W^{\mu\dagger}$) stands for the
$W^{-}$($W^{+}$) field, respectively, and 
$V_{\mu\nu}=\partial_{\mu}V_{\nu}-\partial_{\nu}V_{\mu}$, 
$W_{\mu\nu}=\partial_{\mu}W_{\nu}-\partial_{\nu}W_{\mu}$, 
$\tilde{V}_{\mu\nu}=\frac{1}{2}\varepsilon_{\mu\nu\rho\sigma}V^{\rho\sigma}$. 
The overall normalization is such that the coupling $g_{V}$
is defined as in the SM, $viz.$,
\be
    g_\gamma = e, \qquad g_Z = e \cot \theta_W \ ,
\ee
with $\theta_W$  the weak mixing angle. 
In the SM we have, at the tree level,
\be
\Delta g_1^V = \Delta\kappa_V = \lambda_V = \kt_V 
             = \lt_V = g_4^V = g_5^V = 0 \ 
\ee
Non zero values of the above, usually called anomalous gauge
couplings, would indicate new physics.
The three couplings, $\Delta g_1^V$, $ \Delta\kappa_V $ and $\lambda_V$, 
are even under both $C$ and $P$ transformations. Of the remaining four, 
two $\kt_V $ and $ \lt_V$ violate $P$ but conserve $C$, $g_4^V$ respects $P$ 
but violates $C$, and $g_5^V$ violates both $P$ and $C$.

Eq.~(\ref{lagrangian}) represents the most general $WWV$ Lagrangian
consistent with Lorentz- and gauge-invariance.
Higher derivative terms can be absorbed
into the above couplings provided these are treated as form factors 
and not constants. It is thus important to bear in mind the fact that the
strength of the various terms in the vertex would vary (in general,
independently) with the momentum scales of the process being
considered. The imaginary parts of the form factors are essentially
the absorptive parts of the $WWV$ vertex functions, and,  as such, are
small in the SM or MSSM.  Although absorptive parts that arise
from the same sector of new physics as the anomalous couplings
themselves (as for example in the
models of Ref.~\cite{dynmodel}) need not be suppressed, 
we will assume here that the anomalous couplings are real.

To date, the only direct limits on $CP$-violating $WW\gamma$ couplings 
are \\[1mm]
  ($i$) \mbox{$-0.92<\kt_{\gamma}<0.92$,}\ \mbox{$-0.31<\lt_{\gamma}<0.30$}
        from an analysis of $p\bar{p}\ra W\gamma + X$ events done by the
        D0 Collaboration at Tevatron~\cite{D0:Wgamma}, and 
  \\[1mm]
  ($ii$) $\kt_{\gamma} = 0.11^{+0.71}_{-0.88}\pm 0.09$,
$\lt_{\gamma}=0.19^{+0.28}_{-0.41}\pm0.11$ from the analysis of
$e^+e^- \ra W^+W^-$ and $ We\nu$ data collected by DELPHI
Collaboration~\cite{Delphi}. 

A competitive indirect limit $|\kt_\gamma|<0.6$ has been derived
recently~\cite{gb} from the $b\ra s \gamma$ CLEO data~\cite{cleo},
whereas indirect constraints based on neutron electric dipole moment
(EDM) put a very strong limit $|\kt_\gamma|\lsim 2 \times
10^{-4}$~\cite{edmn}. By investigating $e^+e^-\ra W^+W^-$ at a future
500 GeV linear collider, it has been shown that the constraint
$|\kt_{\gamma}|\leq 0.1$ can be established~\cite{likeli}. Similar
conclusion has been reached for the process $p\bar{p}\ra W\gamma$ at
an upgraded Tevatron~\cite{dawson}. Better limits
of the order of $10^{-3}$ can be reached in the $e\gamma$ and
$\gamma\gamma$ modes of the linear colliders with the polarized
Compton back-scattered photon beams \cite{eg-gg}. As for $WWZ$ couplings, the
reaction $e^+e^-\ra\nu\bar{\nu} Z$ has been examined~\cite{rindani}
with the resulting limit of the order $|g^{Z}_4|< 0.1$. 

Given the tight (but subject to theoretical assumptions)
indirect bound from EDM, it seems 
unlikely that a non-zero $CP$-violating
couplings will be observed directly at future colliders.
Nevertheless, the EDM and the direct observables that we study in this
paper depend on different combinations of the anomalous couplings 
and consequently provide complementary information.
We argue therefore that experimental searches,
wherever possible, should be attempted, if only to overdetermine
the system.

\section{Why $e^+e^- \ra \nu\bar{\nu}\gamma$?}
        \label{sec:nunug}
\begin{figure}[htb]
\input{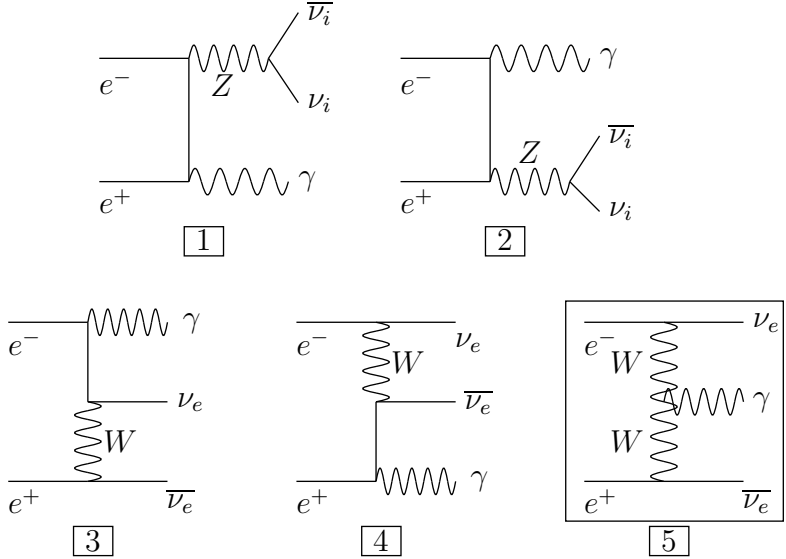}
 \vspace*{2em}
 \caption{\em The Feynman diagrams responsible for
           $e^+ e^- \ra \bar \nu \nu \gamma$.}
      \label{fig:feyn}
\end{figure}

The process $e^+e^-\ra W^+W^-$  necessarily involves both
$WW\gamma$ and $WWZ$ vertices and consequently all 14 couplings. On 
the other hand, the reaction \cite{vvg} 
\be
e^- (p_1) + e^+(p_2) \longrightarrow 
    \nu(p_3) + \bar{\nu}(p_4) + \gamma(p_5),
                       \label{nunug}
\ee
with the Feynman diagrams shown in Fig.~\ref{fig:feyn},  has the
advantage\footnote{The same is also true for both $p\bar{p}\ra W\gamma$ and
                 $e^\pm\gamma \ra W^\pm\nu$.}
 that only the $WW\gamma$ vertex is present. 
Therefore the process with ``a photon + missing
energy'' in $e^+e^-$ annihilation can probe $WW\gamma$ couplings
independently of $WWZ$,  
reducing greatly the number of unknown couplings to be determined 
experimentally.
%
%
Moreover, the number of unknown couplings is further reduced to 
$ \Delta\kappa_\gamma$, $\lambda_\gamma$, $\kt_\gamma$ and  
$\lt_\gamma$, since  
electromagnetic gauge invariance requires that, for on-shell
photons, $\Delta g^\gamma_1 =g^\gamma_4 =g^\gamma_5=0 $, though these
can assume other values for off-shell photons, a fact often missed in
the literature.  

For the process (\ref{nunug}), being formally of 
higher order in the
electroweak coupling than $W$ pair production, one may expect a
reduced sensitivity. However, since the total cross section for this
process increases\footnote{Actually the rapid rise of the cross
section allows for generous kinematical cuts to suppress possible
backgrounds, as we will demonstrate in this paper.} with incoming
energy \cite{jk_dc} until fairly large $\sqrt{s}$, 
while that for $W^+W^-$ decreases strongly with
$\sqrt{s}$, the reduction in sensitivity is less and less severe at
higher energy. Coupled with $W^+W^-$ and $W\gamma$ measurements,
performed at the different momentum transfers, the process
(\ref{nunug}) would allow the possibility of studying the form factor
nature of anomalous couplings.

The sensitivity of $e^+e^-\ra \nu\bar{\nu}\gamma$ to $C$ and $P$
conserving anomalous couplings ($\Delta\kappa_{\gamma}$ and
$\lambda_{\gamma}$) at future linear 
colliders\footnote{First experimental results from this process 
        at LEP2 have been published recently~\protect\cite{lep:photon}.},
has been recently
studied in detail in Ref.~\cite{jk_dc}. Here we will study the
sensitivity to $CP$-violating couplings, namely to $\kt_{\gamma}$ and
$\lt_{\gamma}$. In the static limit they are related to the electric
dipole moment $d_W=e(\kt_\gamma+\lt_\gamma)/2m_W$ and magnetic quadrupole 
moment $Q_W=-e(\kt_\gamma-\lt_\gamma)/m^2_W$ of the $W^+$.

Since, in the process $e^+e^-\ra\nu\bar{\nu}\gamma$, the only
kinematical variables at our disposal are the energy $E_\gamma$ and
the polar angle $\theta_\gamma$ of the produced photon, no truly
$CP$-odd observables can be constructed.  In the absence of the phases of
$\kt_{\gamma}$ and $\lt_\gamma$ we can only look for
the quadratic effects that the $CP$-violating anomalous couplings
induce in the differential distribution of photons. It is therefore
possible to exploit the same $CP$-conserving observables and to follow
the methods of Ref.~\cite{jk_dc} to derive limits on $CP$-violating
photon couplings at future $e^+e^-$ linear colliders.

\section{The SM expectations for $e^+e^-\ra \nu\bar{\nu}\gamma$}
        \label{sec:SM}
The SM cross-section for the process (\ref{nunug})
is best calculated using helicity amplitudes 
and the relevant expressions can be found in Ref.~\cite{AKS}.
Experimentally, the signal comprises a single energetic photon
plus missing momentum carried by invisible neutrinos. The energy and
direction of the photon can be measured with high accuracy. Note
however, that only diagram \mybox{5} of Fig.~\ref{fig:feyn} 
can contribute to the signal, while all diagrams (including \mybox{5})
contribute to the background.
By applying simple kinematical cuts, 
some of these contributions can be suppressed and the
sensitivity to TGC enhanced significantly. 
For example, diagrams \mybox{3} and \mybox{4}
are responsible for an enhancement of the
cross-section at both small photon energy and small emission 
angles. To eliminate these, we impose
\be
25^{\circ}<\theta_{\gamma}<155^{\circ}
        \label{ang_cut}
\ee
as well as\footnote{The same energy cut has been used in the
recent analysis by ALEPH \cite{lep:photon}.} 
\be
E_{\gamma} > 0.1 \sqrt{s}   \ .
        \label{energy_cut}
\ee
Note that cut (\ref{energy_cut}) is different from that of 
Ref.~\cite{jk_dc}, wherein a $\sqrt{s}$--independent cut of
$E_\gamma > 25 \gev$ was imposed. This modification obviously
implies that the selected events
must have higher transverse momentum thus avoiding 
potential background contributions 
from processes such as $e^+e^- \ra \gamma\gamma$ where one
photon disappears down the beam pipe.
This is especially true for larger $\sqrt{s}$.

Similarly, to eliminate events where an 
on-shell $Z$ boson decays to a $\nu\bar{\nu}$ pair 
(diagrams \mybox{1} and \mybox{2} with radiative
$Z$-return), we require here that the photon energy satisfies  
\be
\left| E_\gamma - \frac{s - m_Z^2}{2 \sqrt{s}}
\right| 
     > 5\Gamma_{Z}        \ ,
        \label{minv_cut}
\ee
where $M_{Z}$ and $\Gamma_{Z}$ are the mass and the width of the $Z$
boson. With these cuts, the SM cross section 
(summed over neutrino flavours) is
\be
\Maroon
\sigma_{\rm SM}(\bar \nu \nu \gamma)
        = \left\{
                \barr{ll}
                0.469 \pb & \qquad \sqrt{s} = 350 \gev \\
                0.437 \pb & \qquad \sqrt{s} = 500 \gev \\
                0.361 \pb & \qquad \sqrt{s} = 800 \gev \ .
                \earr
           \right.
                     \label{nngam_cs}
\Black
\ee

\begin{figure}[hb]
\vspace*{-0.0cm}
\hspace*{-0.5cm}
\centerline{
\epsfxsize=7.0cm\epsfysize=6.5cm
\epsfbox{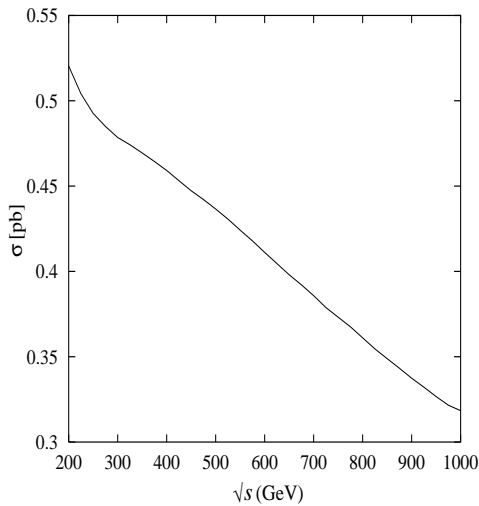} 
}
\caption{\em  The energy dependence of the SM 
  cross section. 
                  }
      \label{fig:rt_s_sm}
\end{figure}

For CM energy in the range (200--1000 GeV), the cross-section falls 
almost linearly (see Fig.~\ref{fig:rt_s_sm}). This is in marked contrast
to Fig.~(2$a$) of Ref.~\cite{jk_dc} where the cross-section was shown to
{\em increase} with $\sqrt{s}$. The difference obviously lies in the 
%
%
stronger form of the energy cut (\ref{energy_cut}) that we use. It 
might seem that the loss of statistics that such a cut entails 
might reduce the sensitivity. However, as we shall show in 
section~\ref{sec:sensit}, this is not really the case.

\section{The anomalous contribution}
        \label{sec:anom_contr}
\begin{figure}[htb]
\vspace*{-0.0cm}
\hspace*{-0.5cm}
\centerline{
\epsfxsize=6.cm\epsfysize=7.0cm
\epsfbox{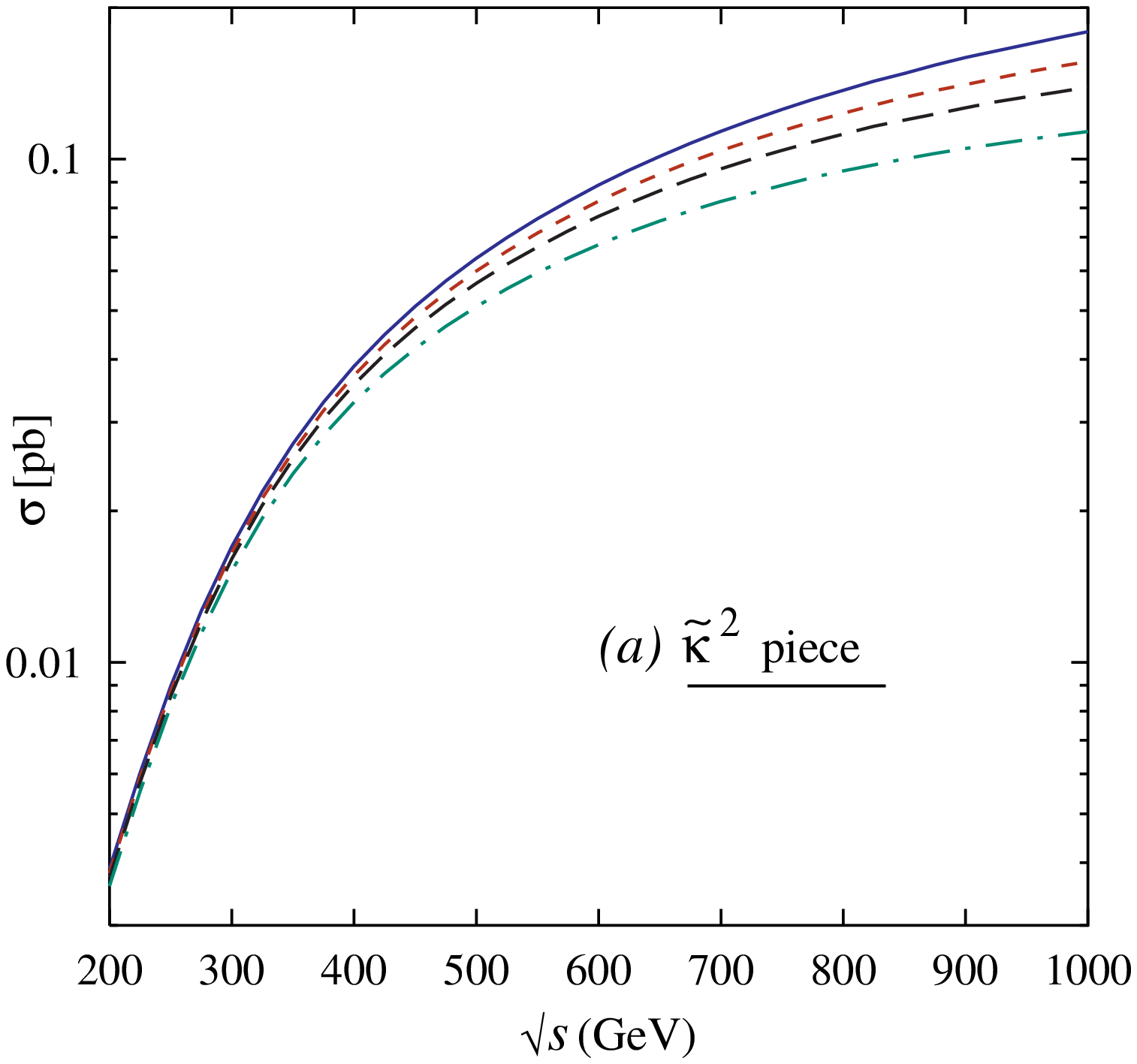} 
\vspace*{-0.0cm}
\hspace*{-1.0cm}
\epsfxsize=6.cm\epsfysize=7.0cm
\epsfbox{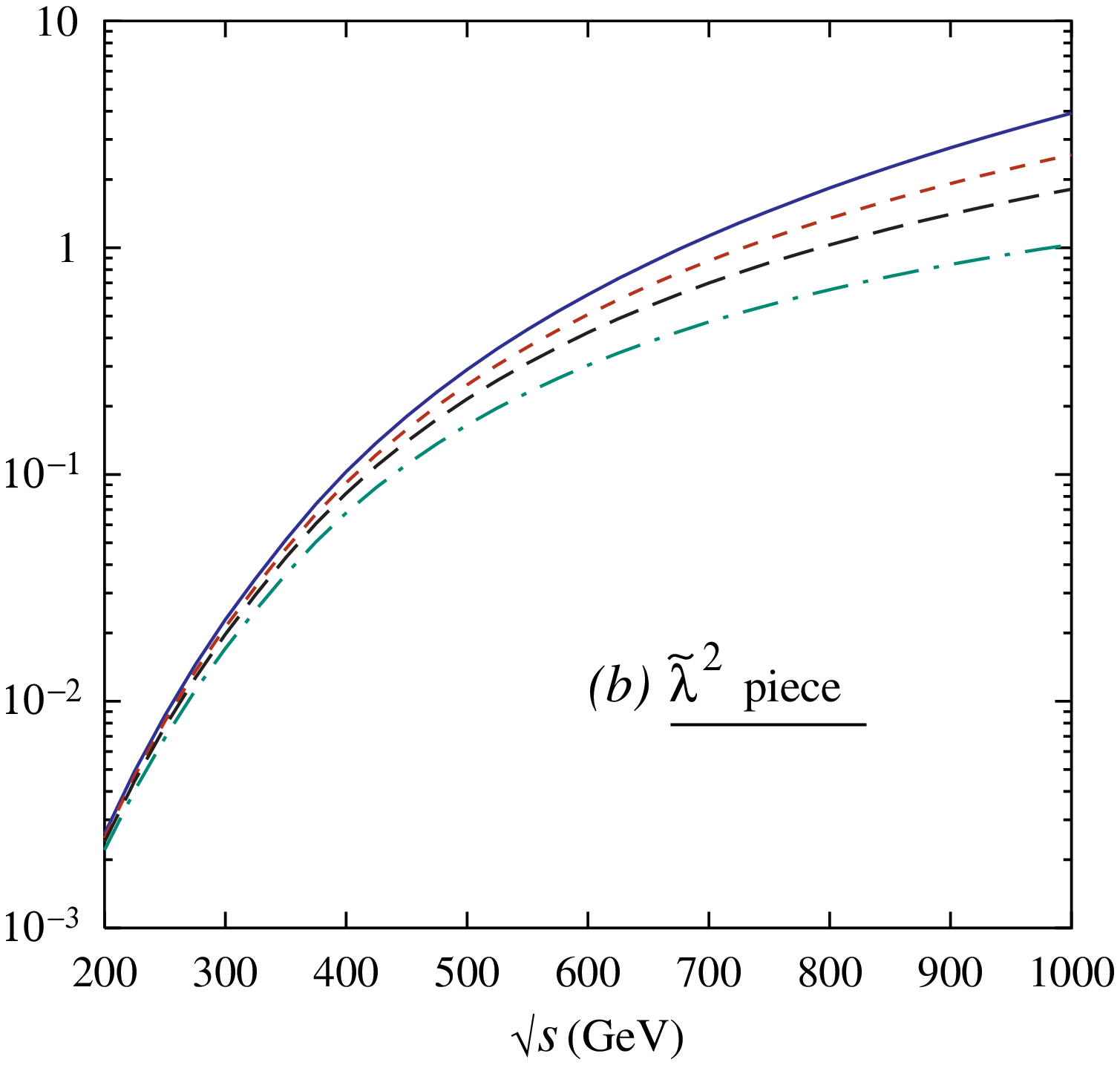}
\vspace*{-0.0cm}
\hspace*{-1.0cm}
\epsfxsize=6.cm\epsfysize=7.0cm
\epsfbox{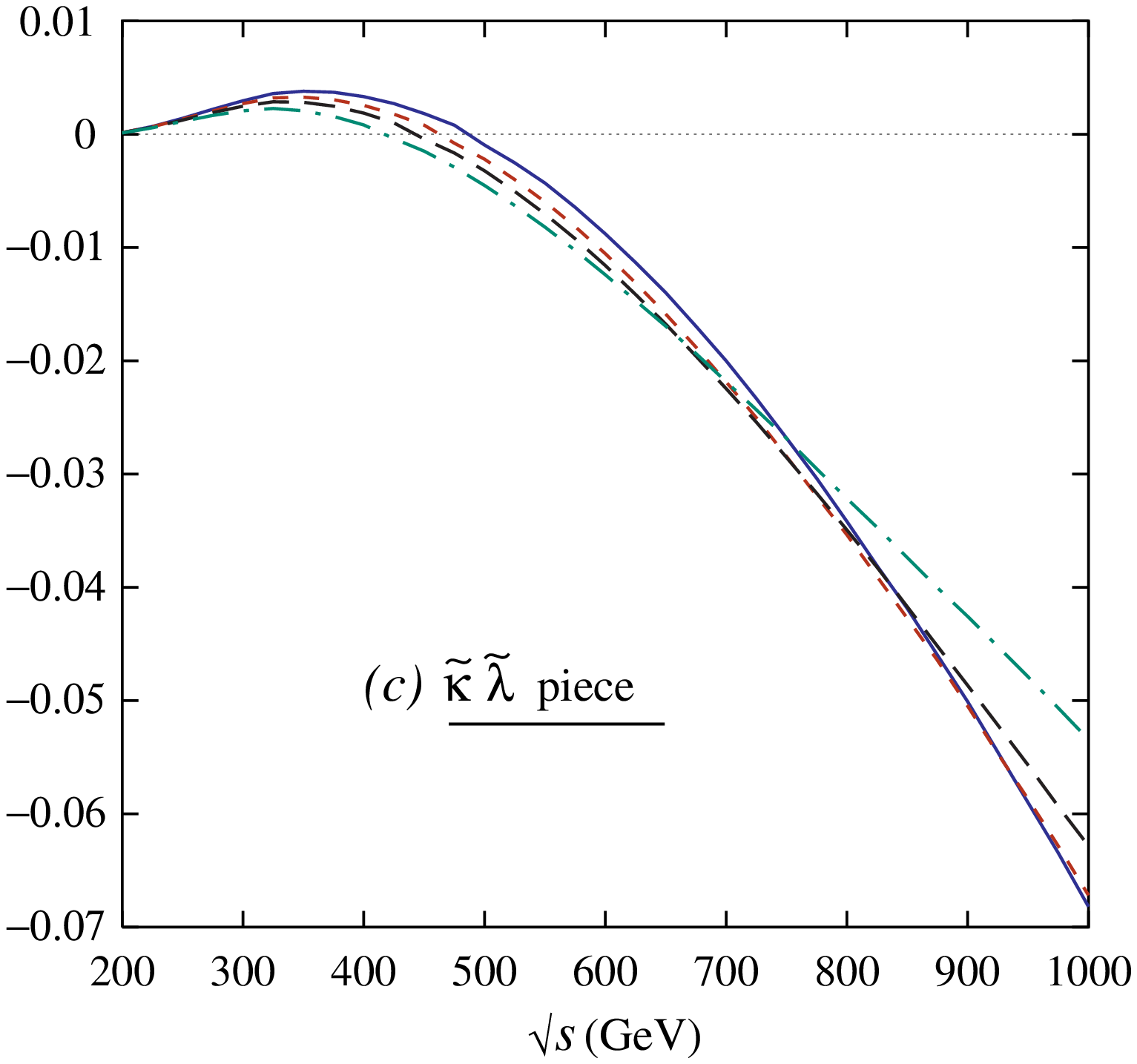}
\vspace*{-0.3cm}
}
\caption{\em The energy dependence of the {\em extra} 
  pieces in the cross section for unit values of the 
  anomalous couplings. The solid, short-dashed, long-dashed and 
  dot-dashed curves correspond to $n = 0, 1, 2, 4$ in 
  eqn.(\protect\ref{unitarity}). The scale ($\Lambda$) of new physics
  has been assumed to be $1 \tev$.
                  }
      \label{fig:rt_s_anom}
\end{figure}

A non-zero value for any one of the anomalous couplings in 
eq.~(\ref{lagrangian}) would imply additional terms in the matrix 
element arising from diagram \mybox{5}. It is easy to see 
that the contributions due to $\tilde{\kappa}_\gamma$ and 
$\tilde{\lambda}_\gamma$ do not interfere 
with the SM amplitude. This is but a reflection of the fact 
that one cannot construct a $CP$-odd observable for the process 
of eq.~(\ref{nunug}). Denoting $d_{ij} \equiv p_i \cdot p_j$, the anomalous
contribution to the spin-summed (averaged) matrix-element-squared 
is then 
\be
\barr{rcl}
\dis \left( \frac{e g^2}{P_{13} P_{24} } \:
              \right)^{-2} \; 
|{\cal M}|^2_{\rm anom}
   & = &  \dis 
           \kt^2 {\cal C}
      + 
         \frac{4 \lt^2}{m_W^4} d_{13} d_{24} 
                 \left\{ {\cal C} + (d_{13} - d_{24})
                                    (d_{15} d_{35} - d_{25} d_{45})
                 \right\}\\[1.5ex]
& + & \dis
         \frac{2 \kt \lt}{m_W^2} 
                 \Big\{ (d_{13} + d_{24}) {\cal C} 
                        + (d_{13} - d_{24}) 
                                    (d_{24} d_{15} d_{35} 
                                        - d_{13} d_{25} d_{45})\\[2ex]
& & \dis \hspace*{3em}
                        - 2 d_{15}  d_{25}  d_{35}  d_{45} 
                 \Big\} \ ,
\earr
        \label{anom_contr}
\ee
where we have suppressed the subscript $\gamma$ in $\kt$ and $\lt$ 
and 
\[
\barr{rcl}
P_{13} & \equiv & 2 d_{13} + m_W^2
        \\
P_{24} & \equiv & 2 d_{24} + m_W^2
        \\
{\cal C} & \equiv & 
        d_{14} d_{25} d_{35} + d_{23} d_{15} d_{45}
     \ .
\earr
\]
The numerical value of the extra contribution can be 
obtained by integrating  $|{\cal M}|^2_{\rm anom}$ 
over the appropriate phase space
volume. In Fig.~\ref{fig:rt_s_anom}, we display this quantity 
as a function of $\sqrt{s}$ for unit values of $\kt$ and $\lt$ 
(and $n=0$; for $n>1$ see below). The 
generalization to arbitrary values is trivial. In contrast to the SM
case (Fig.~\ref{fig:rt_s_sm}), the anomalous contribution {\em grows}
with the CM energy, the effect being more pronounced for dimension 
6 operator, $i.e.$ the non-zero $\lt$ coupling    
(note the different scales on vertical axes).
This is but a consequence of the lack of unitarity for such theories.
%
%

Tree-level unitarity may be restored by postulating a 
form-factor behaviour for $\kt$ and $\lt$. To wit,
\be
    \kt = \kt_0 \left[ \frac{\Lambda^4}
                            {(\Lambda^2 + 2 d_{13}) 
                                (\Lambda^2 + 2 d_{24})}
                \right]^n
    \label{unitarity}
\ee
and similarly for $\lt$~\cite{edmn}.
In eqn.(\ref{unitarity}), $n$ is an integer and 
$\Lambda$ is the scale where new physics manifests itself. 
While $n \geq 1$ ensures that the cross-section falls off 
for sufficiently high $\sqrt{s}$, the effect is not so marked 
for the regime of interest, even for relatively low values 
of $\Lambda$ (see Fig.~\ref{fig:rt_s_anom}). 

A related consequence of this lack of partial wave unitarity  
is that the high-energy end of the photon spectrum gets 
disproportionately populated. In fact, the strongest
dependence on anomalous TGC appears towards the end of the
photon energy spectrum, \ie, in the region which is seriously affected
by the cut~(\ref{minv_cut}) designed to eliminate $\gamma Z$ events. 
Fortunately though, this overpopulation persits for somewhat 
lower values of $E_\gamma$ as well thus allowing us to draw
relatively strong constraints on such couplings.

\section{An estimate of the sensitivity}
        \label{sec:sensit}

In assessing the sensitivity of a future LC we will use the double
differential distribution 
${\rm d}^2 \sigma/{\rm d} E_{\gamma}\,{\rm d} \cos\theta_\gamma $ 
with the phase space divided into a number of bins.  
Choosing a simple $\chi^{2}$ test to
derive 95\% C.L. boundaries in the two-parameter space of
($\kt_{\gamma}$, $\lt_{\gamma}$), we define
\be
\chi^{2}=\sum_{i}^{{\rm \# \ of\ bins}} \left |  
{\frac{N_{SM}(i)-N_{AN}(i)}{\Delta N_{SM}(i)}} \right |^{2} \:,
\label{chiform}
\ee
with $N_{SM}(i)$ and $N_{AN}(i)$ being the 
number 
of events in the bin $i$  expected within the SM and  a theory with the
anomalous TGC, respectively. The error  
$\Delta N_{SM}$ is defined as a combination of statistical and 
systematic errors ({\it cf.}~\cite{jk_dc})
\be
\Delta N_{SM}=\sqrt{(\sqrt{N})^2+(\delta_{syst}N)^{2}} \: .
        \label{delta}
\ee
For our numerical analysis, we use a few 
sets of machine parameters (\ie\  luminosities and CM energies)
considered in the current ECFA-DESY workshop on future
LC~\cite{ecfa-desy}, 
\be
{\cal L} = 
  \left\{
   \barr{rrrc rlll}
     25, &  75, & 100 & \&  & 300 & \fb^{-1} & \qquad \sqrt{s} = 350 \gev \\
     75, & 125, & 300 & \&  & 500 & \fb^{-1} & \qquad \sqrt{s} = 500 \gev \\
    125, & 200, & 500 & \&  & 800 & \fb^{-1} & \qquad \sqrt{s} = 800 \gev \ 
                \earr
           \right. \ .
                     \label{lumin}
\ee
For comparison, we also consider the LEP2 environment with $\sqrt{s}=192$ 
GeV and ${\cal L}=0.5$ and 2 fb$^{-1}$.

We divide the entire range of angular acceptance 
(see eq.~(\ref{ang_cut})) into 26 equal-sized bins of $5^\circ$ each, 
while, for $E_\gamma$, we assume uniform bins of $10 \gev$ each. 
It might seem counterintuitive to use more bins for large $\sqrt{s}$
in view of the smaller SM cross-section. However, this decrease in 
cross-section is more than compensated for by the large increase in the 
proposed luminosity. 
%
\begin{figure}[htb]
\vspace*{-0.0cm}
\hspace*{-0.5cm}
\centerline{
\epsfxsize=8.3cm\epsfysize=8cm
\epsfbox{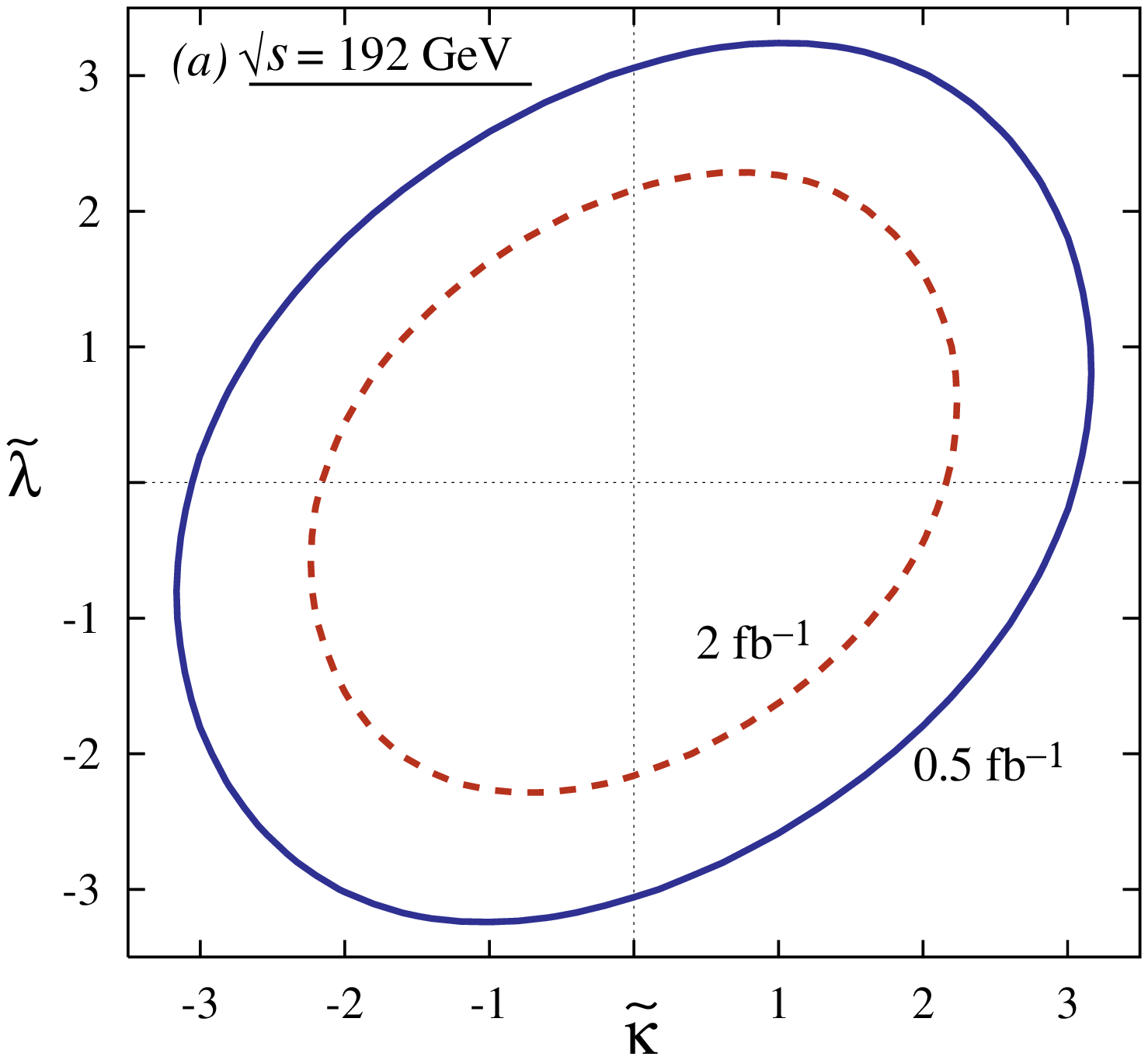} 
\vspace*{-0.0cm}
\hspace*{-0.3cm}
\epsfxsize=8.3cm\epsfysize=8cm
\epsfbox{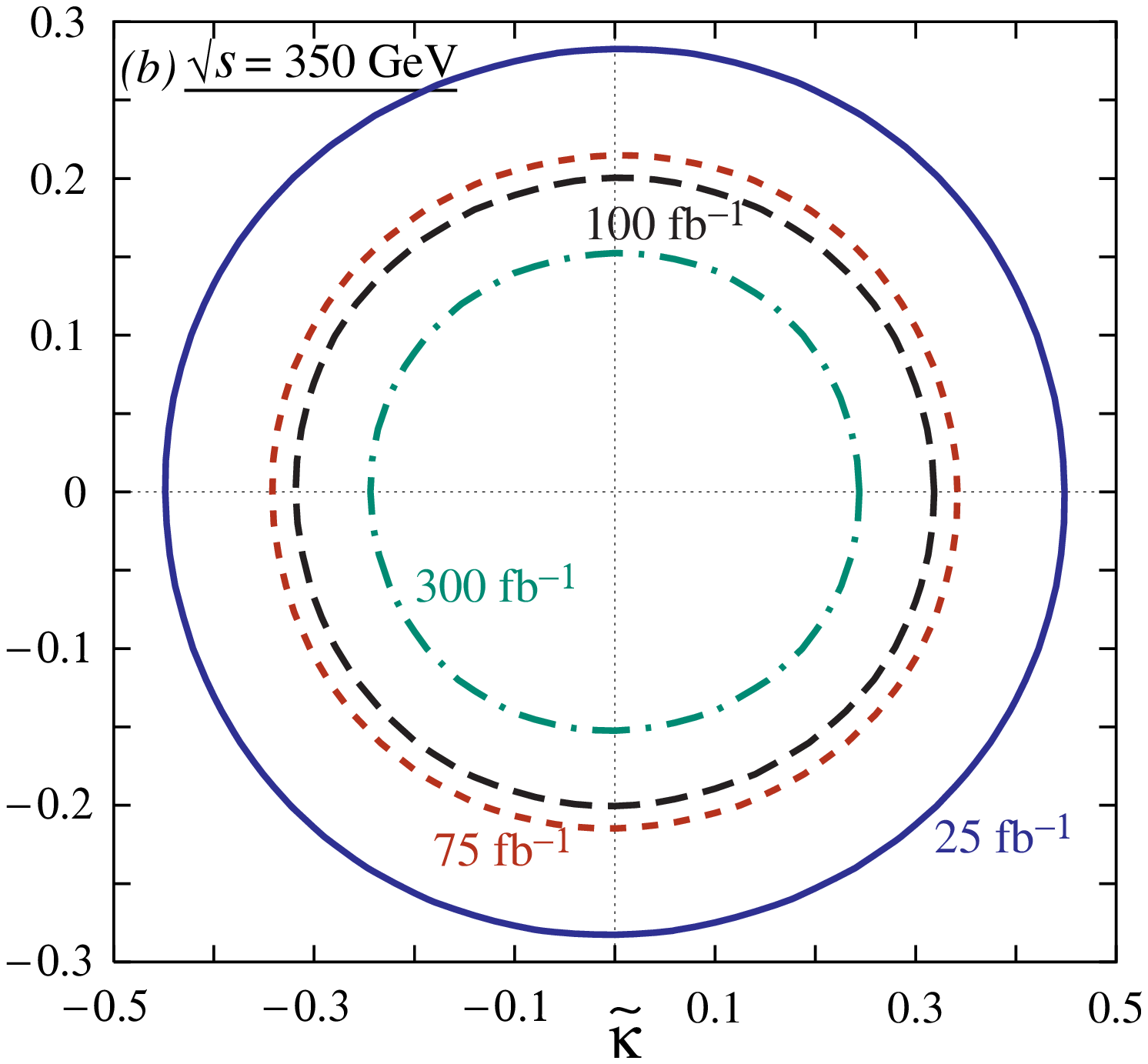}}
\vspace*{-0.5cm}
\centerline{ 
\vspace*{-1cm}
\hspace*{-0.cm}
\epsfxsize=8.3cm\epsfysize=8cm
\epsfbox{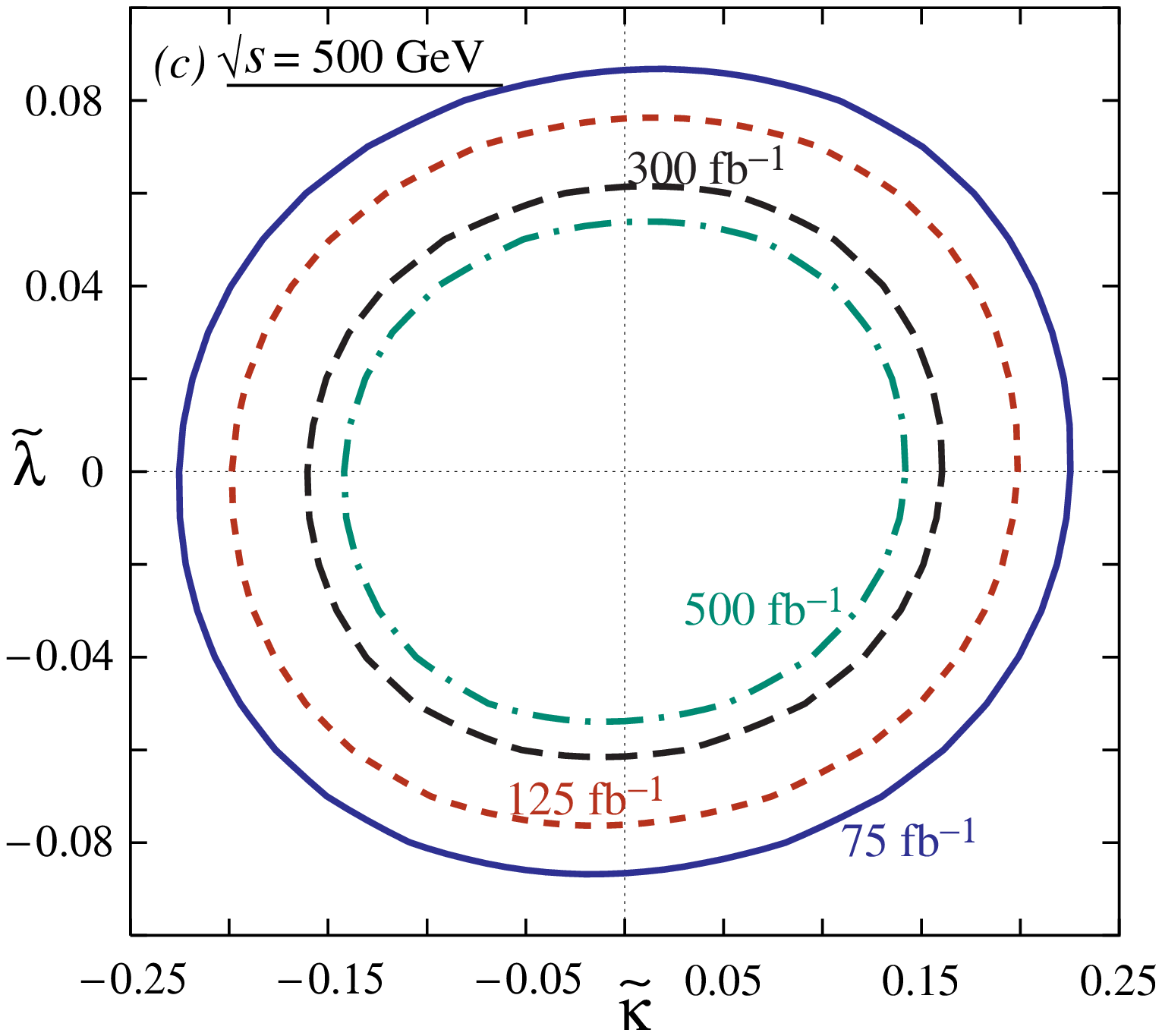}
\vspace*{-0.0cm}
\hspace*{-0.3cm}
\epsfxsize=8.3cm\epsfysize=8cm
\epsfbox{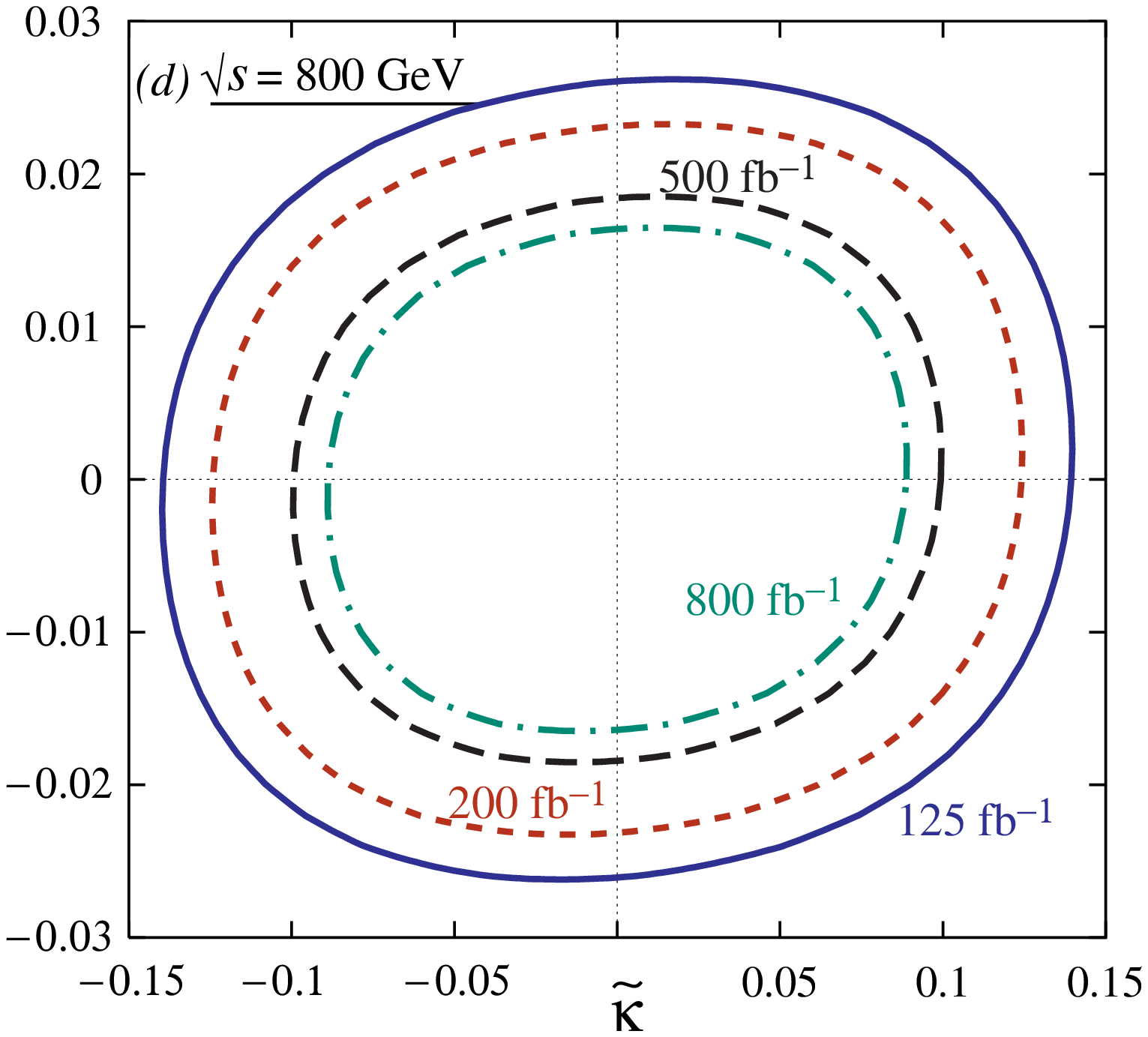}
\vspace*{-2.3cm}
}
\caption{\em 95\% exclusion contours for different 
          energies and luminosities.
         It is assumed that there is no form-factor suppression for the 
         couplings ($n = 0$ or $\Lambda \ra \infty$ in 
         eq.~(\protect\ref{unitarity})).}
      \label{fig:contours}
\end{figure}
%

The systematic error  
$\delta_{syst}$ arises mainly from the detector 
parameters (e.g. uncertainty in the luminosity measurement and 
the detector efficiency). 
We take $\delta_{syst}$ to be 2\%, which is commonly considered as a 
fairly conservative assumption. As it turns out, the error in 
$\Delta N_{SM}$ is dominated by statistical errors and the 
results are insensitive to small changes in $\delta_{syst}$.

For a theory with 2 variables, 95\% C.L. corresponds to 
$\chi^{2}=6$ and the corresponding contours are shown in
Fig.~\ref{fig:contours}. Clearly, a marked improvement
accompanies an increase in luminosity. This reflects the already 
stated fact of $\delta_{syst}$ in eq.~(\ref{delta}) being dominated by
the statistical errors.
Also easy to discern is that a higher CM energy results 
in stronger constraints even for the same luminosity. This is 
expected since such couplings lead to a rapid growth in the 
number of events with $\sqrt{s}$ (see Fig.~\ref{fig:rt_s_anom}).
By the same reasoning, the improvement along the $\lt$ axis 
is more pronounced than that along the $\kt$ axis.

The contours in Fig.~\ref{fig:contours} are nearly elliptical and 
with very little tilt. This of course implies that there is 
only a small correlation between the constraints on $\kt$ and $\lt$,
a feature that we should have expected from a comparison of 
the relative strengths of the $\kt^2$, the $\lt^2$ and the $\kt \lt$ 
pieces in the cross-section (see Fig.~\ref{fig:rt_s_anom}). If one of the 
coupling is known to be identically zero 
(or determined by another experiment),
then the individual bounds on the other can be obtained 
by determining $\chi^2 = 1 $ (68.3\% C.L.) or $\chi^2 = 3.8$ (95\% C.L.)
as the case may be. These bounds 
are summarised in Table~\ref{95results}.
%
%
\begin{table}[htb]
\vspace{-0.5cm}
\begin{center}
\begin{tabular}{||c|c|| l | l||}         
\hline 
\raisebox{-2pt}{$\sqrt{s}$ (GeV)}   &  Luminosity 
		   & \multicolumn{2}{| c ||}{Individual Bounds}
      \\
  \cline{3-4} 
                   & ($\fb^{-1}$) 
                   &  \multicolumn{1}{| c    }{68.3\% C.L.}
                   &  \multicolumn{1}{| c || }{95\% C.L.} 
      \\  
\hline 
 & & & \\[-2.0ex]
192 & 0.5     
    &  $|\kt| < 1.95, \ |\lt| < 1.96$  
    &  $|\kt| < 2.74, \ |\lt| < 2.74$    
	\\[0.7ex]
    & 2     
    &  $|\kt| < 1.38, \ |\lt| < 1.38$
    &  $|\kt| < 1.94, \ |\lt| < 1.94$
	\\\hline 
 & & & \\[-2.0ex]
350 & 75     
    &  $|\kt| < 0.22, \ |\lt| < 0.14$  
    &  $|\kt| < 0.31, \ |\lt| < 0.19$    
	\\[0.7ex]
    & 300
    &  $|\kt| < 0.16, \ |\lt| < 0.098$  
    &  $|\kt| < 0.22, \ |\lt| < 0.14$   
  \\ 
\hline 
 & & & \\[-2.0ex]
500 & 125     
    &  $|\kt| < 0.13, \ |\lt| < 0.049$  
    &  $|\kt| < 0.18, \ |\lt| < 0.069$    
	\\[0.7ex]
    & 500     
    &  $|\kt| < 0.091, \ |\lt| < 0.035$  
    &  $|\kt| < 0.13,  \ |\lt| < 0.049$    
  \\ 
\hline 
 & & & \\[-2.0ex]
800 & 200
    &  $|\kt| < 0.079, \ |\lt| < 0.015$  
    &  $|\kt| < 0.11,  \ |\lt| < 0.021$    
	\\[0.7ex]
    & 800
    &  $|\kt| < 0.057,  \ |\lt| < 0.010$  
    &  $|\kt| < 0.079,  \ |\lt| < 0.015$    
  \\
\hline
\end{tabular}
\end{center}
\vspace{-0.7cm}
\caption{\em Expected individual constraints on $\kt_{\gamma}$ and
       $\lt_{\gamma}$ in the event that the other coupling 
       is already known.}
\label{95results}
\end{table}

Until now, we have sidestepped a few issues, namely the role of 
beam polarization, possible form-factor dependence of the couplings
and the role of the minimum energy cut. The first is easy to estimate. 
Since the signal receives contributions only from left-handed electrons,
polarizing the beam so is expected to help. But this would also increase 
the background by almost as 
much\footnote{Right-handed electrons participate only in diagrams
        \mybox{1} and \mybox{2} of Fig.~\protect\ref{fig:feyn}. But these
        are almost totally eliminated by the cut of 
        eq.~(\protect\ref{minv_cut}).}.
Consequently, the improvement is akin to that resulting from a somewhat
higher luminosity.

\begin{figure}[htb]
\vspace*{-0.0cm}
\hspace*{-0.5cm}
\centerline{
\epsfxsize=7.5cm\epsfysize=7.0cm
\epsfbox{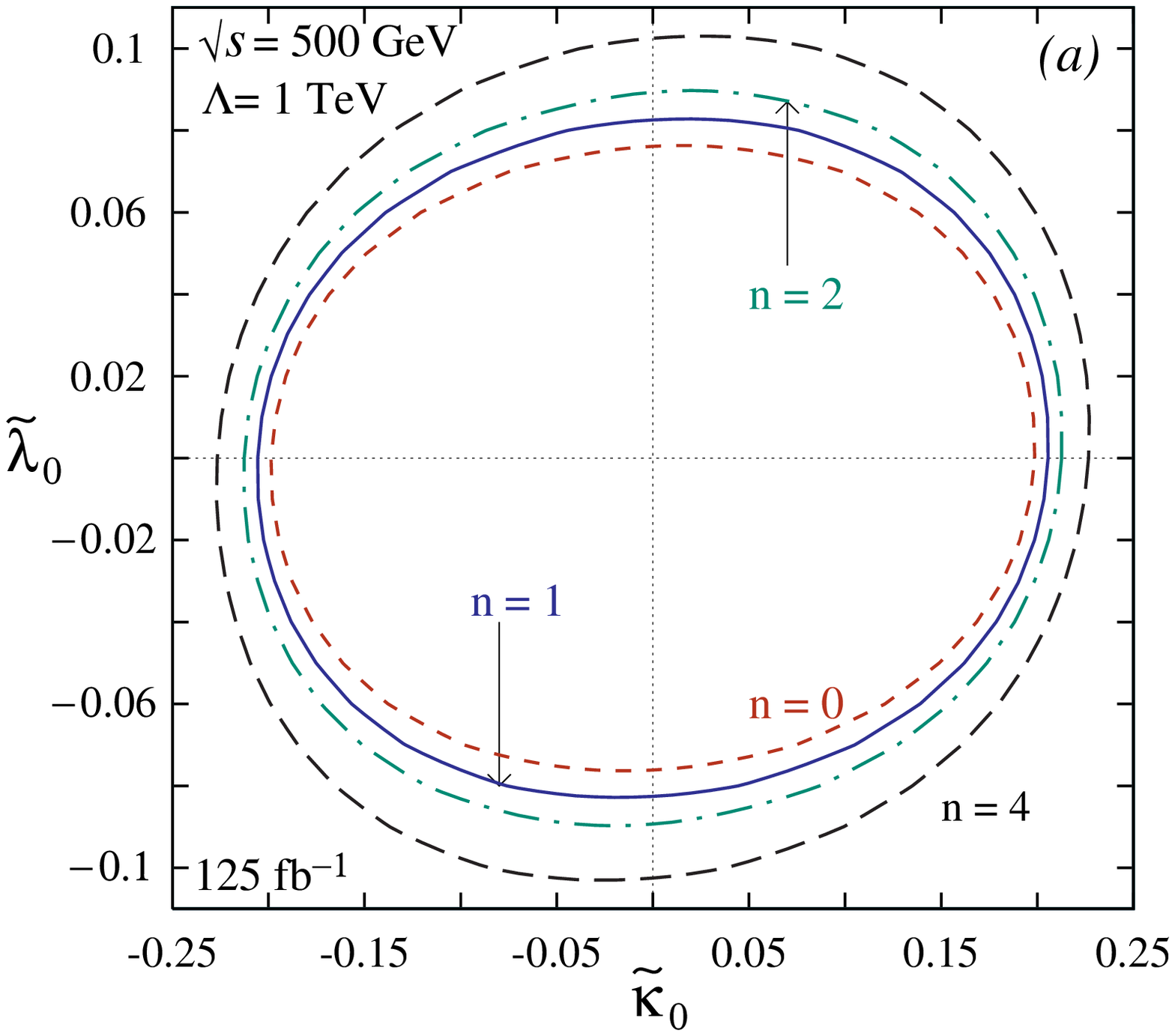} 
\vspace*{-0.0cm}
\hspace*{-0.15cm}
\epsfxsize=7.5cm\epsfysize=7.0cm
\epsfbox{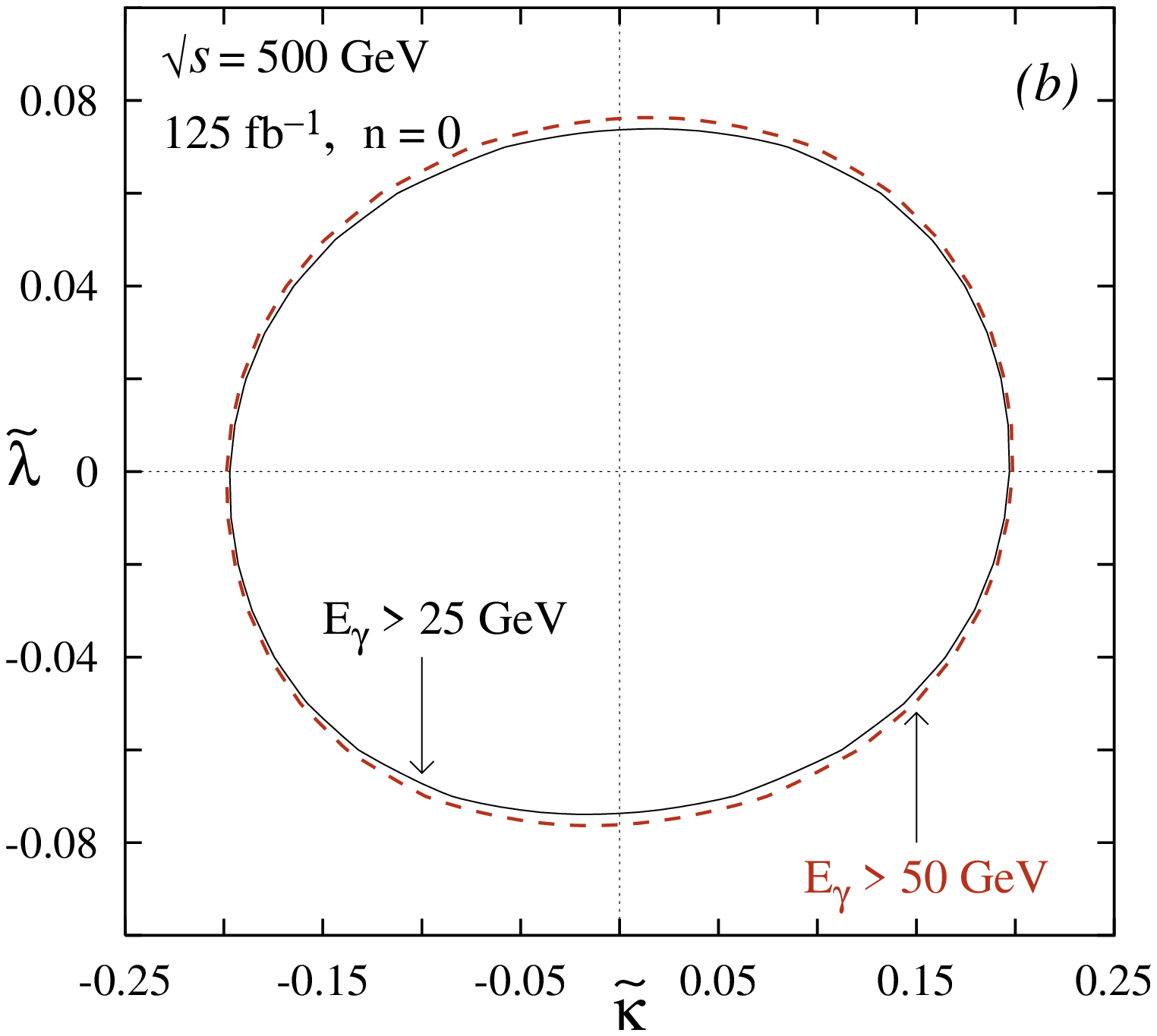}
\vspace*{-0.3cm}
}
\caption{\em 
         (a) The effect of form-factor behaviour 
        (see eq.~(\protect\ref{unitarity})) on the exclusion 
         contour for a given energy and luminosity.
          (b) The effect of the minimum energy requirement
        (see eq.~(\protect\ref{energy_cut})) on the exclusion 
         contour for a given  energy and luminosity. 
         Backgrounds other than those from eq.~(\protect\ref{nunug})
         are deemed to be absent.}
      \label{fig:other_contours}
\end{figure}

As for the form-factor dependence, clearly the anomalous contribution
decreases with increasing $n$ (or smaller $\Lambda$). More importantly,
the high-energy part of the photon spectrum gets depleted faster. This 
implies weaker constraints as evinced by Fig.~\ref{fig:other_contours}($a$).
Again, the effect is more pronounced along the $\lt$ axis. 

Finally, we come to the role of the cut on photon energy 
eq.~(\ref{energy_cut}). As we commented earlier, one role of 
this cut is to eliminate contributions from diagrams \mybox{3} and
\mybox{4} of Fig.~\ref{fig:feyn}. Strictly speaking, this is not 
necessary as the $\chi^2$ function would simply assign a low weight 
to such bins. In fact, in the absence of other backgrounds, 
this cut (as also the other two) only succeeds in rejecting 
additional small but positive contributions to the $\chi^2$. 
Fortunately, this is not a severe loss as Fig.~\ref{fig:other_contours}($b$)
demonstrates clearly. More importantly, eq.~(\ref{energy_cut}) serves 
to eliminate other backgrounds such as $e^+ e^- \ra \gamma \gamma$ 
where one of the photons is missed by the electromagnetic calorimeter.
In a real experimental setup, the precise nature and utility
of such cuts would be dictated by the detector design and hence 
it is premature to dwell on it at further detail.

%
%

\section{Conclusions}  
        \label{sec:concl}

We have investigated the process $e^+e^-\ra\nu\bar{\nu}\gamma$
as a means to derive limits on $CP$-violating couplings
$\kt_\gamma,\:\lt_\gamma$ at a future linear $e^+e^-$ collider. 
Being sensitive only to the $WW\gamma$ couplings, it permits their
independent evaluation. In addition, probing different kinematical
configurations (real photon,
space-like $W$) as compared to that of $e^+e^- \ra W^+W^-$
(real $W$, time-like photon), 
it allows us to probe the form factor behaviour of
the couplings over a distinct region of the momentum transfer. We
have shown that using the differential distribution
$d\sigma/dE_{\gamma}\,d\cos\theta_\gamma$ and appropriate kinematical cuts,
constraints comparable with those expected from $W\gamma$ production at
Tevatron or $W^+W^-$ production at a 500 GeV LC can be obtained. As this
study makes use of all attainable physical information, a
potential improvement of the results can be only achieved by applying
other statistical methods of testing the consistency with the SM.

\newpage
\begin{center}
ACKNOWLEDGMENTS
\end{center}

\noindent AK wishes to thank James Stirling for many 
illuminating discussions.
The early stage of AK's
work was partially supported by the TEMPUS mobility project MJEP 9006.
JK has been partially supported by the Polish Committee for
Scientific Research Grant No. 2 P03B 030 14.


\end{document}